\documentclass[10pt,
  aps,prd,twocolumn,preprintnumbers,superscriptaddress,
  nofootinbib,floatfix]{revtex4-1}
\usepackage[utf8]{inputenc}
\usepackage{amsmath}
\usepackage{bm}
\usepackage{amsfonts}
\usepackage{graphicx}
\usepackage{epstopdf}
\usepackage{hyperref}
\usepackage{array}
\usepackage{soul}
\usepackage{xcolor}
\usepackage{comment}
\usepackage{physics}
\enlargethispage{\baselineskip}
\usepackage{orcidlink}
\usepackage{csquotes}
\hypersetup{
     colorlinks   = true,
     citecolor    = black,
     breaklinks = true,
     urlcolor = black,
     linkcolor = black
}

\newcommand{\article}{{\em article}}
\newcommand{\Appendix}{Appendix}

\newcommand*{\lh}[1]{\lambda_{h}}
 
\newcommand*{\lc}[1]{\lambda_{c}}
\newcommand*{\lo}[1]{\lambda_{0}}

\renewcommand{\vec}[1]{{\bf #1}}
\newcommand{\rmi}[1]{{\mbox{\scriptsize #1}}}
\newcommand{\rmii}[1]{{\mbox{\tiny\rm{#1}}}}
\newcommand{\nn}{\nonumber \\}
\newcommand{\clog}{\Big(c+\ln\frac{3T}{\Lamd}\Big)}
\newcommand{\bmu}{\bar{\mu}}
\newcommand{\Lamd}{\bmu_{\rmii{3d}}}
\newcommand{\LamD}{\bmu}
\newcommand{\gammaE}{{\gamma_\rmii{E}}}
\newcommand{\Veff}{V_{\rmi{eff}}}

\newcommand{\Tc}{T_{\rm c}}

\newcommand{\Tp}{T_{\rm p}}
\newcommand{\Hp}{H_{\rm p}}
\newcommand{\tp}{t_{\rm p}}

\newcommand{\vw}{v_w}

\newcommand{\mD}{m_\rmii{D}}
\newcommand{\yD}{y_\rmii{D}}

\newcommand{\mX}{m_\rmii{$X$}}
\newcommand{\mZ}{m_\rmii{$Z$}}

\newcommand{\Mpl}{M_\rmii{Pl}}

\newcommand{\Lb}{L_b}
\newcommand{\Lf}{L_f}

\newcommand{\re}{\mathop{\mbox{Re}}}

\newcommand\MSbar{$\overline{\rm MS}$}

\newcommand{\UOneCW}{\mathrm{U}(1)_\rmii{CW}}
\newcommand{\SUTwoX}{\mathrm{SU}(2)_\rmii{X}}
\newcommand{\mZprime}{m_\rmii{$Z'$}}
\newcommand{\gCW}{g_\rmii{CW}}
\newcommand{\gX}{g_\rmii{$X$}}

\newcommand{\TQCD}{T_\rmii{QCD}}
\newcommand{\Tshrink}{T_\rmi{shrink}}
\newcommand{\vhqcd}{v_{h,\rmii{QCD}}}
\newcommand{\fpbh}{f_\rmii{PBH}}
\newcommand{\Trh}{T_\rmi{rh}}

\newcommand{\deltac}{\delta_{\rm c}}

\usepackage[normalem]{ulem}

\begin{document}
\title{%
  Thermodynamical uncertainties for primordial black holes\\
  from cosmological phase transitions}

\preprint{SISSA 07/2025/FISI}

\author{Maciej Kierkla\,\orcidlink{0000-0002-2785-5370}\,}
\email{maciej.kierkla@fuw.edu.pl}
\affiliation{Faculty of Physics, University of Warsaw, Pasteura 5, 02-093 Warsaw, Poland}
\author{Nicklas Ramberg\,\orcidlink{0000-0003-1551-9860}\,}
\email{nramberg@sissa.it}
\affiliation{SISSA International School for Advanced Studies, Via Bonomea 265, 34136, Trieste Italy}
\affiliation{INFN Sezione di Trieste, Via Bonomea 265, 34136, Trieste Italy}
\affiliation{IFPU, Institute for Fundamental Physics of the Universe, Via Beirut 2, 34014 Trieste, Italy
}
\author{Philipp Schicho\,\orcidlink{0000-0001-5869-7611}\,}
\email{philipp.schicho@unige.ch}
\affiliation{D\'epartement de Physique Th\'eorique, Universit\'e de Gen\`eve, 24 quai Ernest Ansermet, CH-1211 Gen\`eve 4, Switzerland}
\author{Daniel Schmitt\,\orcidlink{0000-0003-3369-2253}\,}
\email{daniel.schmitt@kit.edu}
\affiliation{Institute for Theoretical Physics, Goethe University, 60438 Frankfurt am Main, Germany}
\date{\today}

\begin{abstract}
\noindent
Strongly supercooled first-order phase transitions
have been proposed as a primordial black hole (PBH) production mechanism.
While previous works rely on simplified models with limited thermodynamic precision,
we stress that reliable theoretical PBH predictions
require precise nucleation dynamics within
realistic extensions of
the Standard Model.
By employing high-temperature dimensional reduction and computing
the one-loop fluctuation determinants,
we provide a state-of-the-art thermodynamic analysis and
obtain an universal lower bound on the transition timescale, $\beta/H_* \simeq 5$.
Then, we
estimate the corresponding PBH abundance for classically conformal gauge-Higgs theories. 
Accounting for constraints from successful percolation and
QCD chiral symmetry breaking,
the parameter space where PBHs are viable dark matter candidates is severely limited.
\end{abstract}

\maketitle

The detection of astrophysical gravitational waves (GWs) by
the LIGO-VIRGO~\cite{LIGOScientific:2016aoc} collaboration
and the first evidence of a stochastic gravitational wave background signal from
pulsar timing array experiments~\cite{NANOGrav:2023ctt,NANOGrav:2023gor,NANOGrav:2023hde,NANOGrav:2023hfp,NANOGrav:2023hvm,NANOGrav:2023icp,NANOGrav:2023tcn,NANOGrav:2023pdq,Antoniadis:2023aac,Antoniadis:2023lym,Antoniadis:2023ott,Antoniadis:2023puu,Antoniadis:2023xlr,Smarra:2023ljf,Manchester:2013ndt,Perera:2019sca,Reardon:2015kba,Manchester:2012za,Kerr:2020qdo} may open a new window into the early Universe.

In the hot primordial plasma, symmetries are restored and eventually broken as the
universe cools, potentially via first-order phase transitions~(FOPTs). In the Standard
Model~(SM), both the electroweak~(EW) and QCD transitions are crossovers%
~\cite{Kajantie:1996mn,Aoki:2006we}. Many beyond the SM~(BSM) scenarios predict a strong
first-order EW phase transition, sourcing GWs in the sensitivity
range of LISA~\cite{2017arXiv170200786A,LISACosmologyWorkingGroup:2022jok}.
A compelling
class of BSM theories are classically conformal extensions of the SM, which yield strong
GW signals due to significant supercooling%
~\cite{Meissner:2006zh,Foot:2007iy,Espinosa:2008kw,Iso:2009ss,Iso:2009nw,Iso:2012jn,
Farzinnia:2013pga,Englert:2013gz,Hashimoto:2013hta,Khoze:2014xha,Hur:2011sv,
Heikinheimo:2013fta,Holthausen:2013ota,Kubo:2014ova,Ametani:2015jla,Kubo:2015joa,
Hatanaka:2016rek,Baratella:2018pxi,Ellis:2018mja,Marzo:2018nov,Ellis:2020nnr,
deBoer:2024jne,Kierkla:2022odc,Sagunski:2023ynd,Kierkla:2023von,Schmitt:2024pby,
Kierkla:2025qyz,Goncalves:2024lrk,Goncalves:2025uwh,Balan:2025uke}.
In slow, strongly supercooled FOPTs, each Hubble patch hosts a few large true vacuum
bubbles, enclosing remnant false vacuum regions. These late-decaying regions induce
large-scale inhomogeneities from statistical fluctuations during nucleation.
If density
perturbations are large enough at horizon re-entry,
they can collapse into primordial
black holes~(PBHs)%
~\cite{Kodama:1982sf,Liu:2021svg,Hashino:2021qoq,Jung:2021mku,Baker:2021nyl,
Baker:2021sno,Baldes:2023rqv,Gouttenoire:2023naa,Gouttenoire:2023pxh,
Gouttenoire:2023bqy,Salvio:2023blb,Salvio:2023ynn,Jinno:2023vnr,Banerjee:2023qya,
Banerjee:2024fam,Flores:2024lng,Ai:2024cka,Conaci:2024tlc,Kanemura:2024pae,
Hashino:2025fse,Balaji:2025tun,Lewicki:2023ioy,Lewicki:2024ghw,Lewicki:2024sfw,
Franciolini:2025ztf,Li:2025nja,Banerjee:2024cwv,Arteaga:2024vde,Goncalves:2024vkj,
Cai:2024nln,Borah:2024lml,Kawana:2022olo}.
The resulting PBH abundance critically depends on the transition timescale.

Previous studies of PBHs induced by supercooled FOPTs rely on
simplified approaches to compute the PT dynamics.
This \article{} presents
a state-of-the-art thermodynamic analysis of
classically conformal SM extensions.
Following~\cite{Kierkla:2025qyz}, we examine strongly supercooled FOPTs  in
the $\UOneCW$ and
$\SUTwoX$ models by
incorporating higher-order thermal corrections through
the three-dimensional effective field theory (3d~EFT) of
high-temperature dimensional reduction~\cite{%
  Ginsparg:1980ef,Appelquist:1981vg,Kajantie:1995dw}.
In flat spacetime, this allows us to compute
the next-to-leading order~(NLO)
thermal bubble nucleation rate~\cite{Hirvonen:2021zej,Gould:2023ovu,Kierkla:2023von} 
including 
fluctuation determinants~\cite{Ekstedt:2021kyx,Ekstedt:2023sqc,Kierkla:2025qyz}.
We find the slowest transitions to have a timescale of $\beta/\Hp = 5.50$ for $\SUTwoX$ and $\beta/\Hp = 5.99$ for $\UOneCW$, respectively. 
Employing the EFT framework enables
a mapping to a wide class of BSM theories,
lending universality to our findings.

Finally, we employ numerical simulations utilizing
{\tt deltaPT}~\cite{deltaPT:2024,Lewicki:2024ghw} and
semi-analytic fitting formulae~\cite{Lewicki:2024sfw,Lewicki:2023ioy,Lewicki:2024ghw}
to estimate the PBH abundance in these models.
We demonstrate that no region of parameter space allows for a sizable PBH abundance,
which is strongly susceptible to the
theoretical thermodynamic uncertainties of supercooled FOPTs.

{\em Model review.}---%
Classically conformal models~\cite{Meissner:2006zh,Foot:2007iy,Espinosa:2008kw,Iso:2009ss,Iso:2009nw,Iso:2012jn,Farzinnia:2013pga,Englert:2013gz,Hashimoto:2013hta,Khoze:2014xha,Hur:2011sv,Heikinheimo:2013fta,Holthausen:2013ota,Kubo:2014ova,Ametani:2015jla,Kubo:2015joa,Hatanaka:2016rek,Baratella:2018pxi,Marzo:2018nov,Ellis:2020nnr,deBoer:2024jne,Kierkla:2022odc,Sagunski:2023ynd,Kierkla:2023von,Schmitt:2024pby,Kierkla:2025qyz,Goncalves:2024lrk,Goncalves:2025uwh,Balan:2025uke}
do not exhibit a mass term in the tree-level potential.
Therefore, the SM Higgs mass squared $\mu_\rmii{SM}^2$
is replaced by a portal coupling to a new scalar field, such that
the potential at tree-level becomes
\begin{equation}
\label{eq:CC_Vtree}
    V_\rmi{tree} =
      \lambda_h H^4
    + \lambda_\phi \Phi^4
    - \lambda_p H^2 \Phi^2
    \,,
\end{equation}
where $H = (G_+, (h+iG_0)/\sqrt{2})$ is the SM Higgs doublet.
The BSM scalar $\Phi$ is charged under a new gauge symmetry, which is spontaneously broken by radiative effects.
As $\Phi$ acquires a vacuum expectation value~(vev),
the EW vacuum is generated via the negative portal coupling in
eq.~\eqref{eq:CC_Vtree},
$\mu_\rmii{SM}^2 = \lambda_p \langle \Phi\rangle^2$.
To keep our findings generic for gauge-Higgs classically conformal theories,
in this \article{},
we consider two such model implementations.


{\em Conformal $\UOneCW$ model.}---%
This model extends the SM with
the $\UOneCW$ Coleman-Weinberg (CW)~\cite{Coleman:1973jx} dark sector.%
\footnote{%
    Conformal U(1) models generically exhibit
    an ultraviolet Landau pole for the coupling constant.
    For certain values of parameters, they can 
    also be stable up to the Planck scale;
    cf.\ e.g.~\cite{Khoze:2014xha,Loebbert:2018xsd}.}
Besides the SM particle content, this model~\cite{Khoze:2014xha} includes a scalar
$\Phi = (\bar{\varphi} + i G)/\sqrt{2}$ charged under $\UOneCW$, and
an Abelian $Z'$ gauge boson with gauge coupling $\gCW$.
We decompose the radial mode as
$\bar{\varphi} = \phi + s$, so that the scalar vev is given by
$
  v_\phi =\mZprime/(Q_\rmii{CW}\gCW)
$
with $Q_\rmii{CW} = 1$ the $\UOneCW$ charge.
In the $\mathrm{U}(1)_\rmii{B-L}$ model, 
e.g., $Q_\rmii{B-L}=2$~\cite{Khoze:2014xha}.
Given input parameters 
$\{\gCW, \mZprime\}$, the quartic $\lambda_\phi$ and portal coupling $\lambda_p$ are 
fixed by requiring the correct EW scale;
see \Appendix{} and~\cite{Kierkla:2022odc,Kierkla:2023von}.
We further set the kinetic mixing parameter to zero at the EW scale, 
$g_\rmi{mix}(\LamD = \mZ) = 0$. Experimental $Z'$ bounds~\cite{ATLAS:2017fih,%
Escudero:2018fwn} imply $\mZprime \gtrsim \mathcal{O}(\mathrm{TeV})$, i.e., 
$v_\phi \gg v_h$ for moderate $\gCW$.

{\em Conformal $\SUTwoX$ model.}---%
This model features a local $\SUTwoX$ symmetry in addition to the SM 
gauge groups~\cite{Hambye:2013dgv,Khoze:2014xha,Prokopec:2018tnq,%
Kierkla:2022odc,Kierkla:2023von,Kierkla:2025qyz}.
The scalar $\Phi = (G_\rmii{X}^+, (\bar{\varphi} + i G_\rmii{X}^0)/\sqrt{2})$, with
$\bar{\varphi} = \phi + s$, is a doublet under $\SUTwoX$ and singlet under the SM.
The $X$ boson mass relates to the zero-temperature vev as
$
  v_\phi = 2 \mX/\gX
$.
For given 
$\{\gX, \mX\}$, the remaining parameters are fixed as in $\UOneCW$.
Reproducing the EW scale requires $\mX \gtrsim \mathcal{O}(\mathrm{TeV})$, so 
$v_\phi \gg v_h$ in both models. Hence, we restrict to tunneling in the 
new-scalar direction~\cite{Prokopec:2018tnq,Kierkla:2022odc}.


To scrutinize the robustness of predictions of PBH abundance,
in this \article{},
we consider the following {\em theoretical uncertainties}:

\hypertarget{sec:Veff:NLO}{(A){\em~Effective action at NLO.}}%
---%
By utilizing the derivative expansion,
we show the form of the effective action of
an arbitrary background field $\phi$.
This is a starting point for calculating the nucleation rate,
and illustrates
the general setup for soft, static, infrared (IR) modes described by
the dimensionally reduced 3d EFT at high temperatures~\cite{Kajantie:1995dw}.

In the soft 3d leading-order~(LO) effective potential,
\begin{align}
\label{eq:Veff:LO:soft}
  V_\rmi{eff}^\rmii{LO}(\phi) &=
    \frac{\mu_{3}^2}{2}\phi^2
  + \frac{\lambda_{\phi,3}}{4}\phi^4
  - \eta_1^{ } \phi^3
  - \bigl(\yD^{ } + \eta_2^{\frac23} \phi^2\bigr)^{\frac32}
  \,,
\end{align}
a barrier is induced by
vector and temporal-scalar ultraviolet (UV) contributions;
see e.g.~\cite{Kierkla:2025qyz,Bernardo:2025vkz}.
The barrier is parametrized by
$\eta_1 \sim g^3$,
the Debye screening mass $\yD \sim \mD^2$, and
the interaction between temporal and Lorentz scalars
$\eta_2 \sim h_3^{3/2} \sim \eta_1$
where in three dimensions
$[\phi] = \frac{1}{2}$,
$[\eta_1] = \frac{3}{2}$,
$[\mD] = [\lambda_{\phi,3}] = 1$.
The effective parameters are listed in the \Appendix.

To compute the LO action,%
\begin{align}
\label{eq:Seff:LO}
  S_{3}^\rmii{LO} = \int_{\vec{x}} \Bigl[
      \frac{1}{2} (\partial_i \phi)^2
      + \Veff^\rmii{LO} (\phi)
    \Bigr]
    \,,
\end{align}
where
$\int_\vec{x} = \int {\rm d}^d \vec{x}$,
we first focus on a simpler setup
and set $\mD \to 0$, yielding
\begin{align}
  V_\rmi{eff}^\rmii{LO}(\phi) &=
    \frac{\mu_{3}^2}{2}\phi^2
  + \frac{\lambda_{\phi,3}}{4}\phi^4
  - \eta\phi^3
  \,,
\end{align}
with $\eta = \eta_1 + \eta_2$,
which isolates 
the soft-enhanced contribution to the barrier~\cite{Bernardo:2025vkz}.
At LO,
the barriers take the following form
\begin{align}
\label{eq:eta_barrier:u1}
  \eta & =
  \frac{g_{\rmii{CW},3}^3}{4\pi}
    \,,
  &&
  \text{for}\quad
  \UOneCW
  \,, \\
\label{eq:eta_barrier:su2}
  \eta & =
  \frac{3g_{\rmii{$X$},3}^3}{32\pi}
    \,,
  &&
  \text{for}\quad
  \SUTwoX
  \,,
\end{align}
revealing that, for fixed model parameters,
the transition weakens with increasing the gauge group rank.

This form of $V_\text{eff}^\text{LO}$ illustrates
  the universality of the EFT approach,
  as different dark sectors yield the same expression with
  appropriately modified 3d couplings.
By utilizing the dimensionless
parameters~\cite{Dine:1992vs,Dine:1992wr,Baacke:1993ne,Baacke:2003uw,Dunne:2005rt,Ekstedt:2021kyx}
$x = \lambda_{\phi,3}/\eta^{2/3}$,
$y = \mu_{\phi,3}^2/\eta^{4/3}$
and $\phi \to \eta^\frac{1}{3} \varphi$,
we can recast the potential in the generic form
\begin{align}
  \widetilde V_\rmi{eff}^\rmii{LO}(\varphi) = 
  \frac{V_\rmi{eff}^\rmii{LO}(\phi)}{\eta^2}
  =
    \frac{y}{2}\varphi^2
  + \frac{x}{4}\varphi^4
  - \varphi^3
  \,.
\end{align}
After redefining
$\vec{x} \to \tilde{
\vec{x}} y^{-\frac{1}{2}} \eta^{-\frac{2}{3}}$ and
$\varphi \to y \varphi$,
the LO action~\eqref{eq:Seff:LO}
takes a dimensionless two-parameter form
\begin{align}
\label{eq:S:LO:fit}
  S_{3}^\rmii{LO} \!&=
  \kappa\!\int_{\tilde{\vec{x}}}\Bigl[
      \frac{1}{2} (\partial_i \varphi)^2
    + \frac{1}{2}\varphi^2
    + \frac{\gamma}{4}\varphi^4
    \!
    - \varphi^3
      \Bigr]
  \equiv \kappa\,\widetilde{S}_{3}(\gamma)
    \,,
\end{align}
where
$\kappa = y^\frac{3}{2}$, and
$\gamma = x y$.
Such a form allows for
fitting the LO 3d action directly~\cite{Matteini:2024xvg, Brdar:2025gyo, Ekstedt:2021kyx, Ekstedt:2022ceo}.

The above procedure extends directly
to computing the action for the full LO soft potential~\eqref{eq:Veff:LO:soft}.
There, we restore a non-vanishing $\mD$ to include
the non-analytic temporal-scalar contribution to the barrier.
At LO, one can either fit the action~\eqref{eq:Seff:LO}
with multiple parameters~\cite{Bernardo:2025vkz}, similar to eq.~\eqref{eq:S:LO:fit},
or compute it numerically~\cite{Kierkla:2023von},
as employed in this~\article{}.

At NLO, the action receives a two-loop correction from the effective potential,
\textit{viz.}
$S_{3}^\rmii{NLO} \supset \int_{\vec{x}} 
      V^\rmii{NLO} (\phi)
$.
The explicit forms of $V^\rmii{NLO}$ for the considered models are found
in~\cite{Hirvonen:2021zej,Kierkla:2023von}
or the \Appendix.

\hypertarget{sec:bubble:nucleation}{(B){\em~Bubble nucleation.}}%
---%
The full thermal bubble nucleation rate
factorizes into a
statistical and dynamical part \cite{Ekstedt:2021kyx, Ekstedt:2022tqk}:
\begin{align}
    \Gamma(T) = A_{\rmi{dyn}} \times  A_{\rm stat}
    \,.
\end{align}
The statistical part incorporates equilibrium effects and can be computed using the 
imaginary-time formalism and 3d EFT. In the saddle-point approximation, 
$A_\text{stat}$ is given by the Euclidean $O(3)$ effective action evaluated on the bounce 
solution $\varphi_b(r)$, where $r$ is the radius of the critical bubble. We compute 
$A_\text{stat}$ up to NLO in the soft expansion following~\cite{Kierkla:2025qyz}.
The LO 
background $\varphi_b$ solves 
$\nabla_r^2 \varphi_b = \pdv{V_3^\text{LO}}{\varphi_b}$ with boundary conditions 
$\varphi_b(\infty) = 0$ and $\partial_r \varphi_b(0) = 0$.

To extend the nucleation rate to full one-loop order, corrections from fluctuations
around the critical bubble $\varphi_b$ must be included. These are encoded in the
fluctuation determinants~\cite{Ekstedt:2021kyx}. For two-loop effects, we employ the
zeroth order of the derivative expansion, i.e.\ the correction to the effective
potential $V_3^\text{NLO}$.
Evaluating the effective action on $\varphi_b$, the
statistical part becomes:
\begin{align}
  A_{\rm stat} =
  \det\nolimits_{\rmii{$S$}}
  \,
  \det\nolimits_{\rmii{$V$}}
  \,
  e^{-S_{\rm 3}^{\rmii{LO}}[\varphi_b] + C[\varphi_b]}
  \,
  e^{-\int_{\vec{r}} V_3^{\rmii{NLO}}[\varphi_b]}
  \,,
\end{align}
where $\det\nolimits_{\rmii{$S$}}$ contains
the one-loop scalar fluctuation determinant, and
$\det\nolimits_{\rmii{$V$}}$ accounts for vector fluctuations.
While non-zero Matsubara modes are integrated out during dimensional reduction
and affect the effective parameters entering into the potential,
only the zero modes remain dynamical and contribute to the fluctuation determinants.
Relevant contributions are discussed
in the \Appendix.
To avoid double counting, the zero-momentum vector mode must be
removed from the LO action~\cite{Gould:2021ccf}.
This is achieved by adding the term
\begin{align}
\label{eq:p0_1loop_gauge_Seff}
    C[\varphi_b] =
  - \frac{1}{12\pi} \int_{\vec{x}} 
  \Bigl[ 
      2n_{\rmii{$V$}}^{ } m_{\rmii{$V$}}^3
    + n_{\rmii{$V$}}^{ } m_{\rmii{$V$,0}}^3
    - n_{\rmii{$V$}}^{ } \mD^3
  \Bigr]
  \,.
\end{align}
Here,
$n_{\rmii{$V$}}$ is the number of gauge modes,
$m_{\rmii{$V$}}$ the mass of  spatial gauge modes,
$m_{\rmii{$V$},0}$ the mass of temporal gauge modes, and
$\mD$  denotes the Debye mass.
For $\SUTwoX$, $n_{\rmii{$V$}}=3$ and for $\UOneCW$, 
$n_{\rmii{$V$}}=1$.

The dynamical part captures non-equilibrium effects of
bubble nucleation~\cite{%
Ekstedt:2022tqk,Gould:2024chm,Hirvonen:2024rfg,Hirvonen:2025hqn},
which we approximate 
without damping as $A_{\rmi{dyn}} \simeq \sqrt{\lambda_-/(2\pi)}$,
where $\lambda_-$ is 
the negative eigenvalue of
the scalar fluctuation operator~\cite{Langer:1969bc,Hanggi:1990zz,Berera:2019uyp,Ekstedt:2023sqc}.
This approximation
often yields a lower bound on
$\beta/H$ in comparison to
the full out-of-equilibrium contributions as seen in e.g.~\cite{Gould:2022ran}.
As a consequence,
the resulting PBH abundance is likely overestimated,
strengthening our conclusion of a negligible PBH abundance.

\hypertarget{sec:percolation:criterion}{(C){\em~Percolation criterion.}}%
---%
Strongly supercooled FOPTs require particular attention during
the computation of the relevant temperature scales.
The PT parameters 
are computed at
the {\em percolation temperature} $\Tp$,
defined via the probability,
$P = \exp(-I(T))$,
for a point in spacetime to remain in
the false vacuum~\cite{1971AdPhy..20..325S,Guth:1979bh,Guth:1981uk,Enqvist:1991xw}.
Utilizing the quantity
\begin{equation}
\label{eq:I(T)}
    I(T) = \frac{4\pi}{3}
    \int_T^{\Tc} \frac{{\rm d} T'}{T'^4} \frac{\Gamma(T')}{H(T')}
    \biggl(\int_T^{T'}\!{\rm d}\widetilde{T} \frac{\vw}{H(\widetilde{T})}\biggr)^3
    \,,
\end{equation}
the moment of percolation, when the PT completes,
is determined by imposing that a certain fraction of the Universe has been converted to
the true vacuum.%
\footnote{%
  In eq.~\eqref{eq:I(T)},
  we fix the bubble wall velocity to $\vw = 1$,
  as we investigate very strong transitions which often
  correspond to detonations.
  For a general perturbative determination of $\vw$,
  see~\cite{Ekstedt:2024fyq}.

}
The typically employed values
\begin{align}
  \label{eq:percolation_criteria}
    I(T_{{\rm p},1}) &= 1
    \, , &
    \mathrm{and} &&
    I(T_{{\rm p},2}) &= 0.34
    \, ,
\end{align}
correspond to
$P_1 \approx 0.34$ and
$P_2 \approx 0.71$.
To additionally avoid eternal inflation~\cite{Guth:2007ng,Borde:1993xh,Guth:1980zm},
we verify that the false vacuum fraction of the entire spacetime volume
decreases around $\Tp$~\cite{Turner:1992tz},
\begin{equation}
\label{eq:condition_false_vacuum_decreasing}
    \frac{1}{V_\mathrm{false}} \frac{\mathrm{d} V_\mathrm{false}}{\mathrm{d}t} =
    H(T) \left(3 + T \frac{\mathrm{d}I(T)}{\mathrm{d}T}\right)\biggr|_{T=\Tp} < 0
    \, .
\end{equation}
If this is not the case, the transition cannot finish successfully at
the estimated $\Tp$.
This is particularly relevant for slow,
or conversely strong PTs.
In parts of the parameter space, the false vacuum volume starts decreasing at a temperature
$\Tshrink < \Tp$, which still allows for a successful transition below $\Tshrink$.
In such cases, it remains unclear at which temperature
the transition parameters should be evaluated~\cite{Ellis:2018mja, Athron:2022mmm, Athron:2023xlk}.

We investigate the regime of extreme supercooling,
where the Universe enters a period of thermal inflation at
the temperature $T_\rmii{V}$, defined by
\begin{equation}
    \frac{\pi^2}{30} g_\epsilon T_\rmii{V}^4 = \Delta \Veff(T_\rmii{V})
    \, ,
\end{equation}
where $\Delta \Veff(T_\rmii{V})$ is the false vacuum energy, and
$g_\epsilon$ denotes the relativistic degrees of freedom related to the energy density.
Since $\Tp \ll T_\rmii{V}$ in the entire parameter space we consider, the Hubble parameter in eqs.~\eqref{eq:I(T)} and~\eqref{eq:condition_false_vacuum_decreasing} is well approximated by
\begin{equation}
\label{eq:H_V}
    H(T) = H_\rmii{V}(T) = \left(\frac{\Delta \Veff(T_\rmii{V})}{3\Mpl^2}\right)^\frac{1}{2} \, ,
\end{equation}
with $\Mpl$ being the reduced Planck mass.

This approximation neglects the contribution to radiation energy density from
the propagating bubble walls%
\footnote{Despite having its source from the potential energy, the energy stored in ultrarelativistic walls redshifts as radiation.
}
\cite{Lewicki:2023ioy, Lewicki:2024sfw}.
Accounting for it requires iteratively solving eq.~\eqref{eq:I(T)}
coupled to the Friedmann equation,
typically lowering the Hubble rate at $\Tp$ and
increasing $\beta/\Hp$,
thereby suppressing PBH formation.
However, since
the no-percolation region from eq.~\eqref{eq:condition_false_vacuum_decreasing} also shrinks, 
this mainly induces an overall shift in parameter space. 
We thus adopt the simplified approach of eq.~\eqref{eq:H_V}.


\hypertarget{sec:qcd:sourced:transitions}{(D){\em~QCD-sourced transitions.}}%
---%
In a large part of the parameter spaces of both models, the conformal PT is sufficiently supercooled to delay EW symmetry breaking to temperatures below the QCD scale.
Then, the cosmic QCD transition occurs first with six massless flavors around
$\TQCD \approx 100\,\mathrm{MeV}$~\cite{Braun:2006jd}.
The breaking of the chiral symmetry through the formation of quark condensates
induces a QCD-scale vev for the SM Higgs~\cite{vonHarling:2017yew,Iso:2017uuu},
$\vhqcd \approx 100$~MeV.
This breaks classical scale invariance (CSI) and induces an additional contribution to the effective potential via the portal term in eq.~\eqref{eq:CC_Vtree},
\begin{equation}\label{eq:QCD_contribution}
  V_\rmi{tree} \supset V_\rmii{QCD} = -\frac{\lambda_p}{4} \vhqcd^2 \varphi^2
  \, .
\end{equation}
Since this negative mass term counteracts the thermal barrier,
it accelerates the conformal PT~\cite{Sagunski:2023ynd,Schmitt:2024pby}.
As a consequence, the inverse timescale $\beta/\Hp$ is enhanced.

{\em Primordial black hole abundance.}---%
Black holes form when an energy overdensity is confined within
a spacetime region smaller than its Schwarzschild radius. 
We examine the impact of thermodynamical uncertainties \hyperlink{sec:Veff:NLO}{(A)}--\hyperlink{sec:qcd:sourced:transitions}{(D)} on PBH formation
during strongly supercooled FOPTs~\cite{Kodama:1982sf,Liu:2021svg,Hashino:2021qoq,Jung:2021mku,Baker:2021nyl,Baker:2021sno,Baldes:2023rqv,Gouttenoire:2023naa,Gouttenoire:2023pxh,Gouttenoire:2023bqy,Salvio:2023blb,Salvio:2023ynn,Jinno:2023vnr,Banerjee:2023qya,Banerjee:2024fam,Flores:2024lng,Ai:2024cka,Conaci:2024tlc,Kanemura:2024pae,Hashino:2025fse,Balaji:2025tun,Lewicki:2023ioy, Lewicki:2024ghw, Lewicki:2024sfw, Franciolini:2025ztf,Li:2025nja,Banerjee:2024cwv,Arteaga:2024vde,Goncalves:2024vkj,Cai:2024nln,Borah:2024lml,Kawana:2022olo}.

PBHs form when the density contrast in a Hubble patch exceeds a critical threshold,
\begin{equation}
\deltac = \frac{\rho(t_k) - \bar{\rho}(t)}{\bar{\rho}(t)}\,,
\end{equation}
where
$\rho(t_k) = \rho_r(t_k) + \rho_v(t_k)$,
$\bar{\rho}(t) = \rho_r(t) + \rho_v(t)$ and $t_{k}$ denotes the time at which we evaluate the density contrast when the mode $k$ re-enters the horizon. 
The threshold $\deltac$ is determined from numerical relativity using
the compaction function at Hubble crossing~\cite{Escriva:2019phb}.
In a radiation-dominated Universe, $\deltac \approx 0.55$~\cite{Franciolini:2022tfm}.

PBH formation requires sufficiently large late-blooming regions of false vacuum, screened by surrounding regions of true vacuum, to collapse~\cite{Gouttenoire:2023naa}.
As pointed out in~\cite{Flores:2024lng}, an overdensity within a false vacuum region is not enough and one should consider the curvature perturbation that renders the universe locally non-flat.
To estimate the PBH abundance, one considers
the mass fraction at formation using the density contrast distribution
$P(\deltac)$ evaluated at
$t = t_k \geq \tp$ where $\tp$ is the time of percolation:
\begin{align}
  \beta_k(M) &= \int_{\deltac}^\infty {\rm d}\delta\, \frac{M}{M_k} P_k(\delta)\,
  \delta_\rmii{D}\!\left(\ln{\frac{M}{M(\delta)}}\right)
  \nn &=
  \frac{\kappa}{\gamma} \left( \frac{M}{\kappa M_k} \right)^{\frac{\gamma + 1}{\gamma}} P_k(\delta(M))
  \,.
\end{align}
Here, $M_k$ denotes the mass within a Hubble horizon,
$\delta_\rmii{D}$ is the Dirac $\delta$-function,
$M$ is the PBH mass, and
$\delta(M) = \deltac + (M/(\kappa/M_k))^{1/\gamma}$.
For a radiation fluid,
$\gamma = 0.38$ and $\kappa = 4.2$~\cite{Franciolini:2022tfm}, provided one considers a scale-invariant curvature power spectrum where $\gamma$ depends on the equation of state, and $\kappa$ depends on the initial profile of the curvature perturbation.
The present-day PBH mass function is
\begin{equation}
  \psi(M) = \int {\rm d}\ln{k}\; \beta_k(M)
  \frac{\rho_r(T_k)}{\rho_{\rm c}}
  \frac{s(T_0)}{s(T_k)}
  \,,
\end{equation}
where
the integration is over all $k$-modes close to the inverse Hubble horizon.
Here, $\rho_{\rm c}$ is the critical energy density of the Universe,
$T_k$ denotes the temperature when the mode $k$ re-enters the horizon, and
$s(T)$ is the entropy density with $T_0$ being the present CMB temperature.
The total PBH abundance normalized to the dark matter (DM) energy density then reads
\begin{align}
    f_\rmii{PBH} &= \frac{\Omega_\rmii{PBH}}{\Omega_\rmii{DM}}
    \, ,&
    \mathrm{with}\quad \Omega_\rmii{PBH} &= \int \mathrm{d} \ln M\; \psi(M)
    \,,
\end{align}
where the integration range is determined
by the initial mass within the horizon~\eqref{eq:H_V} and
its subsequent evolution given the transition timescale.
The PBH fraction of DM is well described by
the fit~\cite{Lewicki:2024sfw,Lewicki:2023ioy,Lewicki:2024ghw}
\begin{equation}
\label{eq:fpbh}
  \fpbh = b_1 \exp\bigl(-b_2 e^{b_3 (\beta/\Hp)}\bigr) \frac{\Trh}{\mathrm{GeV}}
  \,,
\end{equation}
where $\Trh$ is defined in eq.~\eqref{eq:T_rh}. 

For $\beta/\Hp \sim 7.5$,
the correct DM abundance is obtained when
$b_1 \approx 5.5 \times 10^6$,
$b_2 \approx 0.064$, and
$b_3 \approx 0.806$~\cite{Lewicki:2024ghw}.
However, it was recently shown that these values severely overestimate
the resulting PBH abundance~\cite{Franciolini:2025ztf}.
Since $b_3$ impacts the abundance double-exponentially (cf.\ \cite{Wu:2024lrp} for a similar argument),
one can rescale it and instead explore a different target $\beta/\Hp$; see the results below.


The PBH abundance depends on
the post-transition reheating temperature $\Trh$,
computed assuming instantaneous reheating by equating
\begin{equation}
\label{eq:T_rh}
    \Delta \Veff(\Tp) = \frac{\pi^2}{30} g_{\epsilon,\rmi{rh}}^{ } \Trh^4 \, ,
\end{equation}
where
$g_{\epsilon,\rmi{rh}}$ denotes the relativistic degrees of freedom in the broken phase, and
$\Delta \Veff(\Tp)$ is the false vacuum energy released during the transition.

\begin{figure*}
  \centering
  \includegraphics[width=\linewidth]{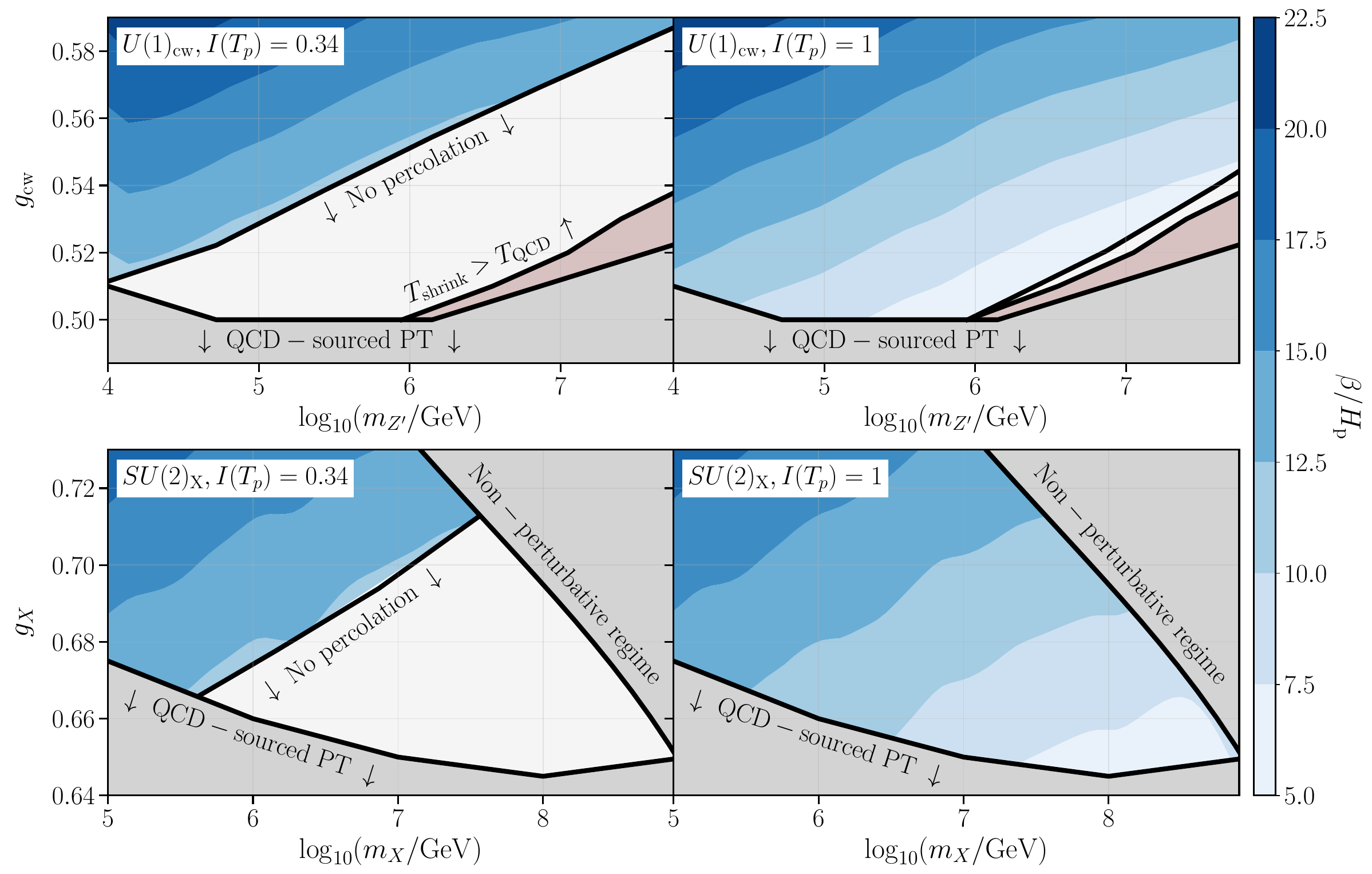}
  \caption{%
    Contour plots of the inverse transition timescale
    $\beta/\Hp$ in
    the $\UOneCW$ and $\SUTwoX$ models,
    shown as a function of
    the gauge boson masses $\mX$, $\mZprime$ and
    couplings $\gX$, $g_\rmii{CW}$ at the input scale defined by the gauge boson masses.
    The values $I(\Tp)$ for the percolation criteria~\eqref{eq:percolation_criteria}
    are indicated, and
    the no-percolation region is shown in light gray.
    Gray regions mark where the PT completes only via
    QCD catalysis or enters a non-perturbative regime for $\SUTwoX$.
    The gray-red area corresponds to $\Tshrink < \TQCD$.
    The oscillations in the contours
    are an artifact of finite grid size in the parameter space scans.
  }
  \label{fig:PBH_abundance}
\end{figure*}

{\em Results.}---%
Figure~\ref{fig:PBH_abundance} shows scans in
$\mZprime$--$\gCW$ space for $\UOneCW$  and
$\mX$--$\gX$ for $\SUTwoX$.
Rows correspond to models, columns to the chosen percolation criterion~\hyperlink{sec:percolation:criterion}{(C)}. 
Color indicates $\beta/\Hp$, with gray regions showing QCD-sourced transitions at small gauge couplings. 
Light gray areas mark unsuccessful percolation at $\Tp$,
where the false vacuum fraction fails to decrease;
see eqs.~\eqref{eq:I(T)} and \eqref{eq:condition_false_vacuum_decreasing}.
In $\SUTwoX$, we exclude regions with non-perturbative $\gX$ near $\Tp$~\cite{Kierkla:2022odc}.


We find minimum inverse timescales $\beta/\Hp = 5.99$~($\UOneCW$) and $\beta/\Hp = 5.50$~($\SUTwoX$), respectively, which increases with decreasing
gauge boson mass or increasing gauge coupling. Achieving a given $\beta/\Hp$ requires
larger $\gX$ in $\SUTwoX$ than $\gCW$ in $\UOneCW$, as seen from
eqs.~\eqref{eq:eta_barrier:u1} and~\eqref{eq:eta_barrier:su2}, which relate the barrier
height to gauge group rank at fixed temperature.

For $I(T_{{\rm p},2}) = 0.34$~(left panels), the minimum inverse timescale we find is $\beta/\Hp \simeq \mathcal{O}(10)$, as benchmark points predicting slower transitions lie deep in the no-percolation region where the standard formalism for estimating $\Tp$~(cf.\ uncertainty~\hyperlink{sec:percolation:criterion}{(C)}) breaks down.
Instead, the transition is expected to finish below $\Tshrink < \Tp$,
where the false vacuum fraction
starts to decrease, as dictated by eq.~\eqref{eq:condition_false_vacuum_decreasing}.
If $\Tshrink < \TQCD$, QCD effects accelerate the conformal PT.
Here, chiral symmetry breaking induces CSI breaking in the $\phi$ potential via
the QCD-scale Higgs vev~\eqref{eq:QCD_contribution}.
The negative portal term counteracts the thermal barrier, facilitating tunneling.
This accelerates the PT, reflected in a steep drop of the bounce action $S_3$ or,
equivalently, an enhanced inverse timescale $\beta/\Hp$~\cite{Ellis:2020nnr,Schmitt:2024pby}.
Thus, the minimum $\beta/\Hp \simeq 5$ obtained using $I(T_{{\rm p},1}) = 1$~represents a universal lower bound.


To our knowledge, all previous PBH-related studies relied on simplified approaches to compute the
thermal nucleation rate. The most common is ``Daisy resummation'' (cf.\ e.g.\
\cite{Gouttenoire:2023pxh, Salvio:2023ynn, Arteaga:2024vde}), which typically
underestimates $\beta/\Hp$ by $\mathcal{O}(50\%)$ compared to the methods used in this
\article{}~\cite{Kierkla:2023von, Kierkla:2025qyz, Croon:2020cgk}. 
To reduce large uncertainties from missing higher-order corrections and large logarithms
in the effective potential, employing the 3d EFT is essential. It is also crucial to
minimize uncertainties in the nucleation rate computation, such as the
breakdown of the derivative expansion~\cite{Kierkla:2025qyz, Gould:2021ccf}.

Employing the fit~\eqref{eq:fpbh}, we identify a narrow region for
$I(T_{{\rm p},1}) = 1$~where PBHs can constitute all of DM. 
However, a recent gauge-covariant study~\cite{Franciolini:2025ztf} found that this
approach significantly overestimates the PBH production rate.
A sizable abundance would thus require even smaller values $\beta/\Hp \ll 4$.
 Such a scenario can be estimated by rescaling $b_3 \to \chi b_3$ in eq.~\eqref{eq:fpbh}. 
Adjusting $\chi$ such that, e.g., $\fpbh(\beta/\Hp = 3) = 1$, we find $\fpbh = 0$ for our most optimist scenario $\beta/\Hp \simeq 5$. 
We have explicitly verified this by using
the updated version of \texttt{deltaPT}~\cite{Lewicki:2024ghw}.

Having investigated two concrete models by employing state-of-the-art thermodynamic methods and estimating the PBH abundance using the updated overdensity criterion~\cite{Franciolini:2025ztf},
we find no parameter space where PBHs are viable DM candidates.
Note that~\cite{Franciolini:2025ztf} neglects, e.g., the energy flux carried by the bubble walls. 
Therefore, the required $\beta/\Hp$ to achieve a sizable PBH population may change in upcoming studies.
We stress that any new threshold value, however, should be above $\beta/\Hp \simeq 5$, which is a universal lower limit allowed by thermodynamics.

{\em Summary.}---%
Theoretical uncertainties in computations of strongly supercooled FOPTs impact
PBH formation as follows:
\begin{itemize}
  \item
    Using state-of-the-art thermodynamics, we find the slowest transitions in the
    $\UOneCW$ and $\SUTwoX$ models yield 
    $\beta/\Hp = 5.99$ and $\beta/\Hp = 5.50$, respectively.
    Successful percolation depends sensitively on the criterion used;
    in particular, criterion $I(T_{{\rm p},2})$ predicts a large no-percolation region.

\item
    For benchmark points with incomplete percolation at $\Tp$,
    the transition terminates at $\Tshrink < \Tp$.
    In $\UOneCW$, this may precede the cosmic QCD transition, while in $\SUTwoX$,
    $\Tshrink < \TQCD$ always.
    QCD-sourced transitions always enhance $\beta/\Hp$ due to CSI breaking by chiral condensates.


\item
  A sizable PBH abundance  requires $\beta/\Hp \lesssim 7.5$ using the criterion of~\cite{Lewicki:2024ghw}, and
  is only possible if
  the percolation criterion $I(T_{{\rm p},1})$ is satisfied.
  With the updated criterion~\cite{Franciolini:2025ztf},
  we do not find the required $\beta/\Hp \ll 4$ anywhere in the viable parameter space.
\end{itemize}

Since small values of $\beta/\Hp$ imply bubble sizes comparable to the Hubble radius, future work will include effects of spacetime curvature on bubble nucleation and growth, also incorporation of the energy transfer of bubbles
modifying the transition timescale~\cite{Coleman:1980aw,Giombi:2023jqq}.
Classically conformal models may also feature
a prolonged matter-dominated era~\cite{Ellis:2020nnr},
which impacts PBH formation.

\begin{acknowledgments}
{\em Acknowledgments.}---%
The authors thank 
Bogumi{\l}a {\'S}wie{\.z}ewska, 
Marek Lewicki, 
Piotr Toczek,
Tuomas V.I. Tenkanen, and
Jorinde van de Vis
for fruitful discussions and comments on the manuscript.
The authors also thank
George Parker for illuminating discussions regarding the notion of fine failing.
DS acknowledges support by the Deutsche Forschungsgemeinschaft (DFG, German Research Foundation) through the CRC-TR 211 'Strong-interaction matter under extreme conditions'– project number 315477589 – TRR 211.
NR acknowledges support in part by INFN TASP.
PS was supported by
the Swiss National Science Foundation (SNSF) under grant
\href{https://data.snf.ch/grants/grant/215997}{\tt PZ00P2-215997}.
MK and this work was supported by Polish National Science Centre under grants 2023/49/B/ST2/02782 and 2023/50/E/ST2/00177.
NR, PS, and DS acknowledge the hospitality of the University of Warsaw during the final stages of this work. MK acknowledges the hospitality of SISSA during the finalization of this work.
\end{acknowledgments}

{\bf Data availability statement:}
The data that support the findings of this article are openly available~\cite{data_FIG1_bib}.

\section*{\Appendix}

\subsection{Matching relations for 3d EFTs}
\label{sec:matching:DRalgo}
This section reports the corresponding matching relations
for the
$\UOneCW$ and
$\SUTwoX$ conformal theories with a portal coupling to the SM.
For the $\UOneCW$ model,
this effectively extends the relations of~\cite{Hirvonen:2021zej}.
The corresponding relations were both computed
using in-house {\tt FORM}~\cite{Ruijl:2017dtg} diagrammatic software,
as well as
{\tt DRalgo}~\cite{Ekstedt:2022bff}.

\subsubsection{Conformal U(1) model}
\label{sec:DR:U1}

The renormalization group equations listed below
are associated with the parameters of
the $\UOneCW$ model
and
encode their running with respect to
the four-dimensional \MSbar{} renormalization scale
$\LamD$ via the $\beta$-functions.
To this end, we use
\begin{equation}
\label{eq:rge:param}
t \equiv \ln\bar{\mu}^2
\;,
\end{equation}
and find
at one-loop level
\begin{align}
\label{eq:beta:gd}
\partial_{t}^{ }\,
\gCW^2 &=
  \frac{1}{(4\pi)^2} \frac{\gCW^{4}}{3}
  \;,
  \\
\label{eq:beta:lambda}\,
\partial_{t}^{ }
\lambda_{\phi}^{ } &=
  \frac{1}{(4\pi)^2} \Big(
      10 \lambda_{\phi}^{2}
    - 6 \gCW^{2} \lambda_{\phi}^{ }
    + 3 \gCW^{4}
  \Big)
  \;.
\end{align}

The matching relations in
the non-conformal limit are given in~\cite{Hirvonen:2021zej}
and are computed here for the conformal $\UOneCW$ model
\begin{align}
g_{\rmii{CW},3}^{2} &=
  \gCW^2T\Big[
    1
    - \frac{\gCW^2}{(4\pi)^2}\frac{\Lb}{3}
\Big]
\;,\\
\label{eq:mD}
\mD^{2} &=
  \frac{\gCW^{2} T^{2}}{3}\Bigl[
      1
    - \frac{{1}}{(4\pi)^2} \Bigl(
        \frac{\Lb - 7}{3} \gCW^{2}
      - 4 \lambda_{\phi}^{ }
    \Bigr)
  \Bigr]
\;,
\end{align}
where
we define a shorthand notation
\begin{align}
\Lb &\equiv
    2 \ln\frac{\LamD}{T}
  - 2 \big( \ln(4\pi) - \gammaE \big)
  \;, \quad
\Lf \equiv \Lb + 4\ln2
    \;, \\
c &=
  \frac{1}{2} \Bigl(
    \ln \frac{8\pi}{9}
    + (\ln\zeta_{2})'
  - 2 \gammaE \Bigr)
  \;,
\end{align}
with
$\gammaE$ being the Euler-Mascheroni constant,
$\zeta_{s}=\zeta(s)$ for $\re\,(s) > 1$ the Riemann zeta function, and
$(\ln \zeta_s)'=\zeta'(s)/\zeta(s)$.
The thermal mass of the complex
${\rm U(1)}$ singlet and
its quartic couplings
take the form
\begin{align}
\label{eq:m:U1:2l}
\mu_{3}^{2} &=
    \frac{T^2}{12} \Big(
      4\lambda_{\phi}
    + 3 \gCW^{2}
  \Big)
    - \frac{T^{2}}{(4\pi)^{2}}\frac{\gCW^{2}}{9}\Big(
        2\gCW^{2}
      - 6 \lambda_{\phi}^{ }
    \Big)
    \\ &
    - \frac{T^{2}}{(4\pi)^{2}}\Lb\Big(
        \frac{13}{12} \gCW^{4}
      - 2 \gCW^{2} \lambda_{\phi}^{ }
      + \frac{10}{3} \lambda_{\phi}^{2}
      \Big)
    \nn &
    - \clog\frac{
        4 g_{\rmii{CW},3}^{4}
      - 8 g_{\rmii{CW},3}^{2} \lambda_{\phi,3}^{ }
      + 8 \lambda_{\phi,3}^{2}
      + h_{3}^{2}
    }{(4\pi)^{2}}
    \;, \nn[2mm]
\lambda_{\phi,3} &= T \Big[ \lambda_{\phi} + \frac{
      \big(2- 3 \Lb\big) \gCW^{4}
    + \Lb \big(
          6 \gCW^{2} \lambda_{\phi}^{ }
        - 10 \lambda_{\phi}^2
      \big)
    }{(4\pi)^2}
    \Big]
  \;, \\[2mm]
h_{3} &= 2\gCW^{2} T \Big[
  1
  - \frac{1}{(4\pi)^2}\Bigl(
        \frac{\Lb-4}{3}\gCW^{2}
      - 8\lambda_{\phi}^{ }
    \Big)
  \Big]
  \;, \\[2mm]
\kappa_{3} &=
  16 \frac{\gCW^{4}T}{(4\pi)^{2}}
\;.
\end{align}
The corresponding effective potential at
the nucleation scale after
integrating out simultaneously
$\mZprime \sim g_{\rmii{CW},3}^{ }\phi_3$ and
$m_\rmii{$Z^\prime _0$} \sim g_{\rmii{CW},3}^{ }T$
is given by
\begin{align}
  \Veff^\rmii{LO} &=
    \frac{1}{2} \mu_{3}^2 \phi_3^2
    + \frac{1}{4} \lambda_{\phi,3}^{ } \phi_3^4
  - \frac{1}{12\pi}\Bigr(
      2m_\rmii{$Z^\prime$}^3
    +  m_\rmii{$Z^\prime_0$}^3
  \Bigr)
  \,,
\end{align}
where
the corresponding mass eigenvalues in the 3d EFT are given by
\begin{align}
  m_\rmii{$H$}^{2} &=
    \mu_{3}^{2} + 3\lambda_{\phi,3} \phi_{3}^2
  \,,&
  m_\rmii{$G$}^{2} &=
    \mu_{3}^{2} + \lambda_{\phi,3} \phi_{3}^2
  \,,
  \nn[2mm]
  m_\rmii{$Z^\prime$}^{2} &=
    g_{\rmii{CW},3}^2 \phi_3^2
  \,,&
  m_\rmii{$Z^\prime_0$}^{2} &=
    \mD^2 + \frac{1}{2} h_{3}^{ }\phi_3^2
  \,.
\end{align}
The corresponding NLO effective potential
consists of temporal and vector contributions and
is given by
\begin{align}
  \Veff^\rmii{NLO} &=
  - \frac{1}{(4\pi)^2} \biggl[ h_3
      \frac{m_\rmii{$Z^\prime_0$}^{2}\! - \mD^2}{4} \Bigl(
        1 + \ln \frac{\Lamd^2}{4 m_\rmii{$Z^\prime_0$}^{2}}
      \Bigr)\\ 
      &\hphantom{=}+ g_{\rmii{CW},3}^{2} m_\rmii{$Z^\prime$}^{2} \Bigl(
        1 + \ln \frac{\Lamd^2}{4 m_\rmii{$Z^\prime$}^{2}}
      \Bigr)
      - \frac{\kappa_{3}}{8} m_\rmii{$Z^\prime_0$}^{2} 
    \biggr]
      \,,
      \nonumber
\end{align}
where
$\Lamd$ is the 3d EFT renormalization scale.
In practice, we then subtract the field-independent contributions as
in~\cite{Kierkla:2023von}.
Finally, we report
the field normalization factor, or kinetic term,
for
the nucleating scalar
\begin{align}
\label{eq:Zphi3:su2}
  Z_{\phi_{3}} &=
    \frac{1}{48\pi}
    \Bigl[
    - 22\frac{g_{\rmii{CW},3}}{\phi_{3}}
    + \frac{1}{4}\frac{h_{3}^{2}\phi_{3}^{2}}{m_\rmii{$Z^\prime_0$}^{3}}
  \Bigr]
  \,.
\end{align}

\subsubsection{Conformal SU(2) model}
\label{sec:DR:SU2}

The matching relations for constructing
both the action and the effective potential
of the conformal $\SUTwoX$ can be found in~\cite{Kierkla:2023von,Kierkla:2025qyz}.
Here, we recollect them for compactness.

The one-loop $\beta$-functions are
\begin{align}
\label{eq:beta:gd:su2}
\partial_{t}^{ }\,
\gX^2 &=
  - \frac{\gX^{4}}{(4\pi)^2} \frac{43}{6} 
  \;,
  \\
\label{eq:beta:lambda:su2}\,
\partial_{t}^{ }
\lambda_{\phi}^{ } &=
  \frac{1}{(4\pi)^2} \Big(
      12 \lambda_{\phi}^{2}
    - \frac{9}{2} \gX^{2} \lambda_{\phi}^{ }
    + \frac{9}{16} \gX^{4}
  \Big)
  \;.
\end{align}
The corresponding matching relations are
\begin{align}
g_{\rmii{$X$},3}^{2} &=
  \gX^ 2T\Big[
    1
    + \frac{\gX^2}{(4\pi)^2}\frac{4 + 43\Lb}{6}
\Big]
\;,\\
\label{eq:mD:SU2}
\mD^{2} &=
    \gX^{2} T^{2} \Bigl[
      \frac{5}{6}
    + \frac{{1}}{(4\pi)^2} \Bigl(
        \frac{430\Lb + 207}{72} \gX^{2}
        + \lambda_{\phi}^{ }
    \Bigr)
  \Bigr]
\;.
\end{align}
The thermal mass of the dark
$\SUTwoX$ doublet and
its quartic couplings
take the form%
\footnote{%
  In comparison to~\cite{Kierkla:2023von},
  we display the corrected $\mu_3^2$.
}
\begin{align}
\label{eq:m:SU2:2l}
\mu_{3}^{2} &=
    \frac{T^2}{16} \Big(
      8 \lambda_{\phi}
    + 3 \gX^{2}
  \Big)
    + \frac{T^{2}}{(4\pi)^{2}}\frac{\gX^{2}}{4}\Big(
        \frac{167}{24}\gX^{2}
      + 3 \lambda_{\phi}^{ }
    \Big)
    \\ &
    + \frac{T^{2}}{(4\pi)^{2}}\Lb\Big(
        \frac{17}{16} \gX^{4}
      + \frac{9}{4} \gX^{2} \lambda_{\phi}^{ }
      - 6 \lambda_{\phi}^{2}
      \Big)
    \nn &
    - \clog\frac{1}{(4\pi)^{2}}\Bigl[
        \frac{39}{16} g_{\rmii{$X$},3}^{4}
      + 9 g_{\rmii{$X$},3}^{2} \lambda_{\phi,3}^{ }
      - 12 \lambda_{\phi,3}^{2}
    \nn &
    \hphantom{- \clog\frac{1}{(4\pi)^{2}}\Bigl[}
      + 6 g_{\rmii{$X$},3}^{2} h_{3}^{ }
      - \frac{3}{2} h_{3}^{2}
      \Bigr]
    \;, \nn[2mm]
\lambda_{\phi,3} &= T \Big[ \lambda_{\phi}
  + \frac{3}{(4\pi)^2}\Bigl(
    \frac{2-3\Lb}{16}\gX^{4}
    + \Lb \frac{
          3\gX^{2} \lambda_{\phi}^{ }
        - 8 \lambda_{\phi}^2
      }{2}
    \Bigr)
    \Big]
  \;, \\[2mm]
h_{3} &= \frac{\gX^{2} T}{2} \Big[
  1
  + \frac{1}{(4\pi)^2}\Bigl(
        \frac{43\Lb+51}{12}\gX^{2}
      + 6\lambda_{\phi}^{ }
    \Big)
  \Big]
  \;, \\[2mm]
\kappa_{3} &=
  \frac{34}{3} \frac{\gX^{4}T}{(4\pi)^{2}}
\;.
\end{align}
The corresponding effective potential at
the nucleation scale after
integrating out simultaneously
$m_\rmii{$X$} \sim g_{\rmii{$X$},3}^{ }\phi_3$ and
$m_\rmii{$X_0$} \sim g_{\rmii{$X$},3}^{ }T$
is given by
\begin{align}
  \Veff^\rmii{LO} &=
    \frac{1}{2} \mu_{3}^2 \phi_3^2
    + \frac{1}{4} \lambda_{\phi,3}^{ } \phi_3^4
  - \frac{3}{12\pi}\Bigr(
      2m_\rmii{$X$}^3
    +  m_\rmii{$X_0$}^3
  \Bigr)
  \,,
\end{align}
where
the corresponding mass eigenvalues in the 3d EFT are given by
\begin{align}
  m_\rmii{$H$}^{2} &=
    \mu_{3}^{2} + 3\lambda_{\phi,3} \phi_{3}^2
  \,,&
  m_\rmii{$G$}^{2} &=
    \mu_{3}^{2} + \lambda_{\phi,3} \phi_{3}^2
  \,,
  \nn[2mm]
  m_\rmii{$X$}^{2} &=
  \frac{g_{\rmii{$X$},3}^2 \phi_3^2}{4}
  \,,&
  m_\rmii{$X_0$}^{2} &=
    \mD^2 + \frac{1}{2} h_{3}^{ }\phi_3^2
  \,.
\end{align}
The corresponding NLO effective potential
consists of temporal and vector contributions and is
given by
\begin{align}
  \Veff^\rmii{NLO} &=
    \frac{3}{(4\pi)^2} \biggl[
      - {h_3} \frac{m_\rmii{$X_0$}^{2}\! - \mD^{2}}{4} \Bigl(
        1 + \ln \frac{\Lamd^2}{4 m_\rmii{$X_0$}^{2}}
      \Bigr)
      + \frac{5\kappa_{3}}{8} m_\rmii{$X_0$}^{2}
  \nn &
  \hphantom{{}=\biggl[}
    + g_{\rmii{$X$},3}^{2}\Bigl(
        m_\rmii{$X$}^{ } m_\rmii{$X_0$}^{ }
      + \frac{3}{2} m_\rmii{$X_0$}^{2}
  \\ &
  \hphantom{{}=\biggl[ +g_{\rmii{$X$},3}^{2} \Bigl(}
      + \frac{4m_\rmii{$X_0$}^{2} - m_\rmii{$X$}^{2}}{2}
        \ln \frac{\Lamd^{ }}{m_\rmii{$X$}^{2} + 2 m_\rmii{$X_0$}^{ }}
  \nn &
  \hphantom{{}=\biggl[ + g_{\rmii{$X$},3}^{2}\Bigl(}
      + \frac{m_\rmii{$X$}^{2}}{16}\Bigl(
          7  + 8\ln 2
          + 21\ln \frac{\Lamd^2}{9 m_\rmii{$X$}^{2}}
          \Bigr)
      \Bigr)
    \biggr]
      \,.
      \nonumber
\end{align}
In practice, we then subtract the field-independent contributions as
in~\cite{Kierkla:2023von}.
Finally, we report
the field normalization factor, or kinetic term, for
the nucleating scalar
\begin{align}
\label{eq:Zphi3:u1}
  Z_{\phi_{3}} &=
    \frac{1}{16\pi}
    \Bigl[
    - 11\frac{g_{\rmii{$X$},3}}{\phi_{3}}
    + \frac{1}{4}\frac{h_{3}^{2}\phi_{3}^{2}}{m_\rmii{$X_0$}^{3}}
  \Bigr]
  \,.
\end{align}

\subsection{Fixing input parameters}\label{app:input_params}
In the 3d~EFT, we use
tree-level
vacuum values for the above couplings. 
This procedure can be extended to
one-loop corrected relations,
as described in appendix~A of~\cite{Niemi:2021qvp}
and~\cite{Kajantie:1995dw,Lewicki:2024xan,Bhatnagar:2025jhh}.

The parameters of both classically conformal models are fixed
by the following scheme~(for details in the $\SUTwoX$ model,
see appendix C of \cite{Kierkla:2022odc}):
\begin{itemize}
  \item[\bf In:]
    Gauge coupling
    $\gCW$ for $\UOneCW$ and
    $\gX$ for $\SUTwoX$ as well as
    the corresponding
    pole masses of $\mZprime$ or $\mX$.
    Here, we associate pole masses with \MSbar-masses.
    The parameters are defined at the \MSbar-scale
    $\LamD = \mX$.
  \item[(i)]
    minimize the scalar potential at the loop level
    to determine the new scalar vev; this fixes scalar quartic $\lambda_\phi$,
  \item[(ii)]
    run the parameters to and compute the scalar vev at the \MSbar-scale $\LamD = \mZ$, match the portal coupling $\lambda_p$ to generate the EW vacuum,
  \item[(iii)]
    run the parameters to the matching scale
    $\LamD = X \overline{T}$  with
    $\overline{T}=\pi  T$
    using
    one-loop $\beta$-functions
    \eqref{eq:beta:gd}--\eqref{eq:beta:lambda} and
    \eqref{eq:beta:gd:su2}--\eqref{eq:beta:lambda:su2}.
    Here, $X$ is a constant typically varied to quantify
    the importance of higher-order corrections.
    We set $X=1$ throughout the analysis.
  \item[\bf Out:]
    \MSbar-parameters as function of physical parameters and $\overline{T}$; 3d EFT parameters.
\end{itemize}
Note, the minimization of the loop-corrected potential to
fix the input parameters can be improved by using the full
one-loop vacuum renormalization. 
By relating pole masses to physical two-point functions,
the improved scheme ensures that higher-order corrections in
the renormalization conditions can be included and the momentum dependence of
the pole masses is respected.
See e.g.~\cite{Kajantie:1995dw,Niemi:2021qvp,Lewicki:2024xan}.

\subsection{Fluctuation determinants}
\label{sec:bubble:details}

In our computation of the thermal bubble nucleation rate,
we evaluated the fluctuation determinants for
one-loop corrections from the scalar and gauge modes.
For the $\SUTwoX$ conformal model,
we reuse the results from~\cite{Kierkla:2025qyz}. To compute the fluctuation determinant in the conformal $\UOneCW$ model, we adapt results obtained in~\cite{Ekstedt:2021kyx, Kierkla:2025qyz}.
Here, we briefly summarize the relevant operators.

The scalar-mode determinant containing one-loop scalar mode corrections is given by,
\begin{align}
\label{eq:detS_softEFT}
\det\nolimits_{\rmii{$S$}} &=
\mathcal I_\phi \sqrt{ \biggl|\frac{\det\hphantom{'}\mathcal O_\phi(\varphi_{\rmii{F}})}{\det'\mathcal O_\phi(\varphi_{b})}\biggr|}
  \,, &
  \mathcal I_\phi &= \left(\frac{ S_{\rm 3}^{\rmii{LO}}[\varphi_{b}] }{2\pi}\right)^{3/2}
  \,,
  \nn[2mm]
  \mathcal O_\phi &= - \partial^2 + (\Veff^\rmii{LO})^{\prime\prime}
  \,,
\end{align}
where
$\mathcal I_\phi$ is the Jacobian resulting from going to collective coordinates
(see e.g.~\cite{Ekstedt:2023sqc, Ekstedt:2021kyx}),
$\varphi_{b}$ denotes the LO bounce solution,
while $\varphi_{\rmii{F}}$ is a false vacuum solution,
which in our case is given by $\varphi_{\rmii{F}}=0$.
The primed $\det'$ denotes a determinant without negative eigenvalues.
This contribution can be evaluated numerically using
{\tt BubbleDet}~\cite{Ekstedt:2023sqc}.
The explicit form of the operator
$\mathcal O_\phi$ is model-dependent, as it includes
the second-derivative of the LO effective potential.

For the $\SUTwoX$ model,
the vector determinant contains contributions from
mixed spatial gauge modes and Goldstone $\det\nolimits_{\rmii{$XG$}}$, transverse gauge modes $\det\nolimits_{\rmii{$X_T$}} $,
temporal gauge modes $\det\nolimits_{\rmii{$X_0$}}$,
and ghosts $\det\nolimits_{\rmi{g}}$:
\begin{equation}\label{eq:vecdet}
\det\nolimits_{\rmii{$V$}} =
    \det\nolimits_{\rmii{$X_0$}}
    \det\nolimits_{\rmi{g}}
    \det\nolimits_{\rmii{$X_T$}} 
    \det\nolimits_{\rmii{$XG$}}
  \,.
\end{equation}
For the detailed form of these determinants and
the illustration of their numerical evaluation,
we refer to \cite{Kierkla:2025qyz}.
For the $\UOneCW$ model,
the overall form of the vector determinant is identical, and differs only in the details of
the gauge fluctuation operators that
contain different expressions for
the gauge modes masses.
The gauge determinants can be also evaluated by using
a derivative expansion, {\em viz.}
\begin{align}
    \det\nolimits_{\rmii{$V$}} \simeq C[\varphi_b]
    + \int_{\vec{r}} \frac{Z_{\phi_{3}}[\varphi_b]}{2} (\partial_i \varphi_b)^2
    + \mathcal{O}(\partial^4)
    \,,
\end{align}
where
$C$ is the zero-momentum part of
the one-loop gauge fluctuations, which is a correction to the effective potential and
was introduced in eq.~\eqref{eq:p0_1loop_gauge_Seff}.
The explicit form of $Z_{\phi_{3}}$ can be found in
eqs.~\eqref{eq:Zphi3:u1} and \eqref{eq:Zphi3:su2}.
This expansion behaves well for the temporal gauge modes, as they are massive even
at $\varphi_{\rmii{F}}$.
On the other hand,
the expansion always breaks down for the gauge-Goldstone determinant as
the corresponding modes become inevitably massless
at the tail of the bounce solution~\cite{Kierkla:2025qyz, Gould:2021ccf}. 

\bibliography{ref}

\end{document}